\documentclass[preliminary,copyright,creativecommons]{eptcs}
\providecommand{\event}{TERMGRAPH 2016} 
\usepackage{amsmath,multicol}

\usepackage{pict2e,macros,xspace,amssymb,prooftree,linlog,nets,color} 
\newcommand\Red[1]{#1}

\begin{document}

\title{Compiling Process Networks to Interaction Nets}
\def\titlerunning{Compiling Process Networks to Interaction Nets}

\author{Ian Mackie\institute{LIX, CNRS UMR 7161, \'Ecole
    Polytechnique, 91128 Palaiseau Cedex, France}}

\def\authorrunning{I. Mackie}

\maketitle

\begin{abstract}
Kahn process networks are a model of computation based on a collection
of sequential, deterministic processes that communicate by sending
messages through unbounded channels. They are well suited for
modelling stream-based computations, but are in no way restricted to
this application. Interaction nets are graph rewriting systems that
have many interesting properties for implementation. In this paper we
show how to encode process networks using interaction nets, where we
model both networks and messages in the same framework.
\end{abstract}

\section{Introduction}

We relate two models of computation: Kahn Process Networks (KPNs) and
Interaction Nets. Our aim is to encode process networks as Interaction
Nets so that we can make use of implementations, and also to
understand a different way of programming and computing within
interaction nets.  Specifically, we investigate encoding a KPN as a
system of interaction nets, so from one perspective we can see this
work as giving a compilation. Since there are many implementations of
interaction nets this gives an implementation technique for process
networks.

Kahn Process Networks~\cite{KahnG:semslp} model computation using
tokens travelling around a fixed network, and the structure of the
network does not change during computation. Each node of the network
transforms, depending on its function, tokens by consuming and
creating new tokens. Transformation of tokens happens asynchronously
and in parallel in the network: many tokens can be travelling around
the network, and different nodes can be transforming different tokens
at the same time. The communication channels, where the tokens flow,
are buffered: the order of tokens is preserved, so tokens are queued
waiting to be processed. Each processing element must block if inputs
are not yet available, but they must not block to output. Only one
process can write to each channel, and only one process can read each
channel.

Interaction nets~\cite{LafontY:intn} on the other hand are a specific
form of graph rewriting system. A program is represented as a network,
and the net is rewritten to normal form by graph rewrite rules, called
interaction rules. These rules act locally: no part of the graph can
be copied or erased globally, so consequently one interaction cannot
interfere with another interaction. Interaction rules can be applied
asynchronously and in parallel; a feature shared with process
networks. Each node of an interaction net must be connected by a
single edge port-to-port to another node. Thus there are similarities
between Interaction Nets and KPN.

Our goal is to simulate process networks in interaction nets. We
achieve this by modelling both the fixed network and tokens using
nodes in interaction nets. The firing of a transformation in the
process network becomes an interaction rule (or several interaction
rules).  It is then just a convention that some nodes in the
interaction net are fixed and others are travelling---in reality, we
are just performing a rewrite rule.  Not only does this give an
implementation of data-flow networks with interaction nets, but it
offers an interesting programming style that is frequently more
natural (and often more efficient) than other ways. We see this as an
application of interaction nets where the body of research on parallel
implementation is becoming stronger.

\paragraph{Overview.}

The rest of this paper is structured as follows. In the next section
we give some background material on process networks and set up the
specific style of interaction nets that we are using, including some
notation and conventions that we adopt. In Section~\ref{sec:blocks} we
build some example interaction systems that serve as building blocks
for our compilation of data-flow networks. Section~\ref{sec:compiling}
gives an outline of the compilation process, and gives some examples
of process networks in interaction nets.  Section~\ref{sec:examples}
gives an example of using the compilation, and also gives additional
examples of programming with interaction nets in the style of
process networks. Finally, we conclude in Section~\ref{sec:conc}.

\section{Background}\label{sec:background}

\paragraph{Process networks.}

A Kahn Process Network~\cite{KahnG:semslp,KahnMacQueen} is a directed
graph $G=(V,E)$ with sequential processing units as the vertices $V$
and unbounded buffers as the connecting edges $E$. A process, which is
a sequential program that could be written in any language (for
example C, Java, Algol, Haskell, etc.), communicates with another
process if there is a connecting edge, called a channel, linking the
two processes.  A communication channel (stream or buffer) is a finite
or infinite sequence of data elements (all the examples in this paper
are numbers, but other data, included structured, is possible).  We
use a list notation $S = [x_1,x_2,...]$, where the empty stream is
denoted $[\;]$, when we need to write them down.

A process can be understood as a function mapping a collection of
streams to another collection of streams. The following diagram
illustrates the idea, where there are $m$ inputs and $n$ outputs:

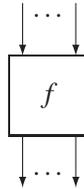
\begin{net}{40}{70}
  \putbox{10}{20}{30}{30}{$f$}
  \putDvector{15}{20}{20}
  \putDvector{35}{20}{20}
  \puttext{25}{5}{$\cdots$}
  \putDvector{15}{70}{20}
  \putDvector{35}{70}{20}
  \puttext{25}{65}{$\cdots$}
\end{net}

Each process $f$ is a sequential program that can read from input
streams $S_1, \ldots, S_m$ to write to output streams $U_1,\ldots,
U_n$ (we will assume $m,n>0$) that can include the two primitives for
communication:

\begin{itemize}
\item \texttt{send x on U}: this command sends a value $x$ (we will
  assume all values are integers in this paper) along the channel
  $U$. The command is non-blocking, which is why the connecting edges
  are unbounded buffers, working as a FIFO queue.

\item \texttt{x = wait(U)}: this command waits for a value on channel
  $U$, and assigns the value to the variable $x$. This command is
  blocking: a read on a channel must wait until the value is there,
  and only then consumes it.
 
\end{itemize}

If a process has only write instructions, then it may be the case that
$m=0$ is required. Such processes are said to be able to process on
empty buffers, and this is typically needed to start the computation
going. Here we will favour an alternative and initialise channels
instead. If a process has repeated processing on empty buffers, then
we can simulate this behaviour by creating a cycle by connecting an
input from an output. The data on this extra channel can then be used
to trigger the process. The process can then read this input, and just
send it back around the loop to the input.  In this way, all processes
trigger output from input, but the price to pay is that we might need
to initialise some of the channels with starting values.

Within a process, the \texttt{send} and \texttt{wait} instructions can
be used in a number of different ways:

 \begin{itemize}
  \item A process does not have to read all of the inputs. Thus some
    channels can accumulate data indefinitely if a process is writing
    to the channel but no process is reading it.
        
  \item The order of input/output is not constrained in any way. A
    process can read from some of the channels in any order, then
    write to others. It can also interleave reading and writing.
 \end{itemize}

A process has local variables, and thus an internal state. This is
used to make the process behave differently on future inputs, and also
to store past results. In the following, interaction nets will use
different principal ports to mimic the different behaviour. The
internal state of a process will be represented by the corresponding
internal state of an interaction net node.

A process network is a collection of processes connected together. The
edges are shared channels where communication can take place.  The
network of processes has a linearity constraint: only one writer per
channel and one receiver per channel. A process can duplicate values
on a channel (and write a value to two different channels) but each
channel must be linear in this sense.  A consequence of these
conditions is that processes are deterministic. But they are also
amenable to a high degree of concurrency: many processes can be active
at one time, and each processor need only wait for all it's values
before proceedings. For this reason process networks are a model of
distributed computation.  There are many variants of this model, for
example synchronous processes. There are also many studies of
properties of these networks, such as deadlock. Here we are interested
in implementing them rather than studying any properties.

We give an example adapted from~\cite{KahnG:semslp}. In this network
$f$ is a process that alternately receives input from the left/right
and copies it to the output channel. $g$ is a process that receives an
input and alternately copies it left and right. $h_i$ is a process
that initially emits an $i$, then copies the input to the output
channel.

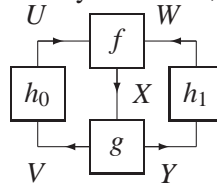
\begin{net}{80}{60}
  \putbox{30}{0}{20}{20}{$g$}
  \putbox{30}{40}{20}{20}{$f$}
  \putbox{0}{20}{20}{20}{$h_0$}
  \putbox{60}{20}{20}{20}{$h_1$}
  \putHline{10}{10}{20}
  \putHline{10}{50}{20}
  \putHline{50}{10}{20}
  \putHline{50}{50}{20}
  \putVline{10}{10}{10}
  \putVline{70}{10}{10}
  \putVline{10}{40}{10}
  \putVline{70}{40}{10}
  \putVline{40}{20}{20}
  \putLvector{30}{10}{10}
  \putRvector{10}{50}{10}
  \putRvector{50}{10}{10}
  \putLvector{70}{50}{10}
  \putDvector{40}{40}{10}
  \puttext{10}{60}{$U$}
  \puttext{60}{60}{$W$}
  \puttext{10}{0}{$V$}
  \puttext{50}{30}{$X$}
  \puttext{60}{0}{$Y$}

\end{net}

The code for these processes is given by the following, where we just
show, respectively, $h_0$ and $f$:

\begin{multicols}{2}\small
\begin{verbatim}
  send 0 on U
  repeat
    x = wait(V)
    send x on U
  end


bool b = true
repeat
  if b then x = wait(U)
       else x = wait(W)
  send x on X
  b = not b
end
\end{verbatim}
\end{multicols}

In this example, if we monitor the output of $f$, we get an infinite
alternating sequence of $0$ and $1$. This process does not deadlock,
and no synchronisation is needed for any of the individual processes.

We refer the reader to the literature (see for
example~\cite{KahnG:semslp}) for a more detailed description of
data-flow networks.  Faustini~\cite{Faustini82} gives an operational
semantics of process networks. There are many works implementing KPN
(see for example \cite{ParksR03} where they are encoded in Java), and
there are many programming languages based on the principles.  We give
examples of these networks using interaction nets later.

\paragraph{Interaction nets.}

Analogous to term rewriting systems, we have a set of user-defined
nodes (drawn as circles and squares in this paper), and a set of rewrite rules:

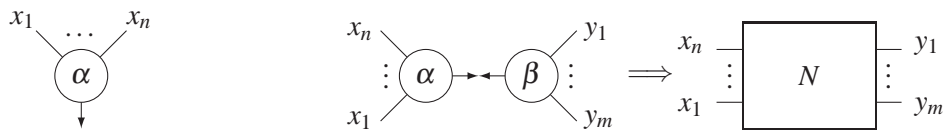
\begin{center}
\begin{mnet}{40}{45}
  \putalpha{20}{20}
  \putDvector{20}{10}{10}
  \putline{12.9}{27.1}{-1}{1}{10}
  \putline{27.1}{27.1}{1}{1}{10}
  \puttext{20}{35}{$\cdots$}
  \put(2.6,38){\makebox(0,0)[br]{$x_1$}}
  \put(37.4,38){\makebox(0,0)[bl]{$x_n$}}
\end{mnet}
\qquad\qquad\qquad\qquad
\begin{mnet}{200}{45}
\putalpha{20}{20}
\putbeta{60}{20}
\putRvector{30}{20}{10}
\putLvector{50}{20}{10}
\putline{12.9}{27.1}{-1}{1}{10}
\putline{12.9}{12.9}{-1}{-1}{10}
\putline{67.1}{27.1}{1}{1}{10}
\putline{67.1}{12.9}{1}{-1}{10}
\puttext{5}{23}{$\vdots$}
\puttext{75}{23}{$\vdots$}
\put(0,0){\makebox(0,0)[br]{$x_1$}}
\put(0,40){\makebox(0,0)[tr]{$x_n$}}
\put(80,0){\makebox(0,0)[bl]{$y_m$}}
\put(80,40){\makebox(0,0)[tl]{$y_1$}}
\puttext{105}{20}{$\Lra$}
\putbox{140}{0}{50}{40}{$N$}
\putHline{130}{10}{10}
\putHline{130}{30}{10}
\puttext{135}{23}{$\vdots$}
\putHline{190}{10}{10}
\putHline{190}{30}{10}
\puttext{195}{23}{$\vdots$}
\put(125,5){\makebox(0,0)[br]{$x_1$}}
\put(125,35){\makebox(0,0)[tr]{$x_n$}}
\put(205,5){\makebox(0,0)[bl]{$y_m$}}
\put(205,35){\makebox(0,0)[tl]{$y_1$}}
\end{mnet}
\end{center}
The set of rewrite rules is constrained: there is at most one rule for
each pair of nodes; and each name in the diagram occurs exactly
twice---once in the left, and once in the right. This means that the
interface is preserved by reduction, and a consequence of this is that
duplication and erasing must be done explicitly, but the rewriting
system is one-step confluent by construction.

We extend interaction nets in two ways, and here we will present this
informally. First, we allow nodes to hold values (just numbers are
needed for the examples in this paper, but other base types can be
included in a similar way).  Consequently, rules can inspect and
update these values. It is possible to do this and still preserve the
same confluence properties by placing conditions on the rules. Next,
we avoid introducing auxiliary nodes that are essentially required
because arguments are taken one at a time (cf.\ Currying). We do this
by allowing some nodes to have several principal ports, and the
rewrite rule requires all to match. This offers no new computational
power, but groups several interactions which is sometimes
convenient. We illustrate all these ideas with an example. Consider
the two rules which encode pairwise addition on lists of numbers:

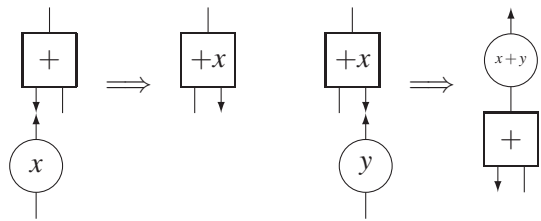
\begin{center}
\begin{mnet}{90}{80}
\putagent{15}{20}{$x$}
\putbox{10}{50}{20}{20}{$+$}
\puttext{50}{50}{$\Lra$}
\putbox{70}{50}{20}{20}{$+x$}
\putVline{20}{70}{10}
\putVline{15}{0}{10}
\putVline{25}{40}{10}
\putVline{75}{40}{10}
\putUvector{15}{30}{10}
\putDvector{15}{50}{10}
\putDvector{85}{50}{10}
\putVline{80}{70}{10}
\end{mnet}
\qquad
\begin{mnet}{90}{80}
\putagent{25}{20}{$y$}
\putbox{10}{50}{20}{20}{$+x$}
\putVline{20}{70}{10}
\putVline{25}{0}{10}
\putVline{15}{40}{10}
\putDvector{25}{50}{10}
\putUvector{25}{30}{10}
\puttext{50}{50}{$\Lra$}
\putUvector{80}{70}{10}
\putagent{80}{60}{{\tiny$x+y$}}
\putbox{70}{20}{20}{20}{$+$}
\putDvector{75}{20}{10}
\putVline{85}{10}{10}
\putVline{80}{40}{10}
\end{mnet}
\end{center}
we can combine these into a single rule in the following way:

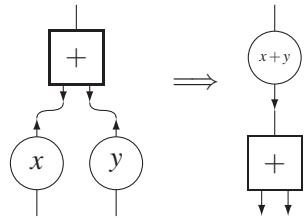
\begin{net}{120}{90}
\putagent{10}{20}{$x$}
\putbox{15}{50}{20}{20}{$+$}
\putagent{40}{20}{$y$}
\puttext{70}{50}{$\Lra$}
\putagent{100}{60}{\tiny $x+y$}
\putVline{25}{70}{10}
\putVline{10}{0}{10}
\putVline{40}{0}{10}
\putDvector{30}{50}{7}
\putUvector{10}{30}{7}
\putUvector{40}{30}{7}
\putDvector{20}{50}{7}
\putbox{90}{10}{20}{20}{$+$}
\putDvector{100}{50}{10}
\putVline{100}{30}{10}
\putDvector{105}{10}{10}
\putDvector{95}{10}{10}
\putVline{100}{70}{10}
\qbezier(10,37)(10,40)(15,40)
\qbezier(15,40)(20,40)(20,43)
\qbezier(40,37)(40,40)(35,40)
\qbezier(35,40)(30,40)(30,43)
\end{net}

For the programs we write, all the properties of interaction nets are
preserved, and this extension can be implemented in usual interaction
nets.  There are many implementations of interaction nets (see for
example \cite{HassanMS15,MackieS15}), including parallel ones. We will be building
interaction net systems to simulate process networks, and for this to
work we assume that the implementation is fair, by which we mean that
no new reductions are done before older ones are completed. Most
implementations respect this, so this avoids any starvation or
live-lock issues that might arise.

\section{Building blocks}\label{sec:blocks}

In addition to the examples for interaction nets given in the last
section, we build some examples here that will be useful later for the
compilation, specifically for the representation of the communication
channels (buffers).

\paragraph{Lists and streams.}
Using two kinds of nodes, we can build lists of numbers:

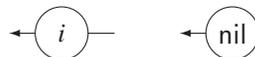
\begin{center}
\begin{mnet}{40}{20}
\putLvector{10}{10}{10}
\putagent{20}{10}{\small $i$}
\putHline{30}{10}{10}
\end{mnet}
\qquad
\begin{mnet}{40}{20}
\putLvector{10}{10}{10}
\putagent{20}{10}{\nil}
\end{mnet}
\end{center}

An example list with four elements would look like this:

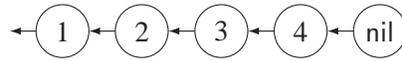
\begin{net}{150}{20}
\putagent{20}{10}{1}
\putagent{50}{10}{2}
\putagent{80}{10}{3}
\putagent{110}{10}{4}
\putagent{140}{10}{\nil}
\putLvector{10}{10}{10}
\putLvector{40}{10}{10}
\putLvector{70}{10}{10}
\putLvector{100}{10}{10}
\putLvector{130}{10}{10}
\end{net}

Streams are just lists without \nil. We can write simple operations
over these lists/streams. For example, applying an operation to each
element of a list (cf.\ map) can be done using the following two
interaction rules, where we increment each element of the list:

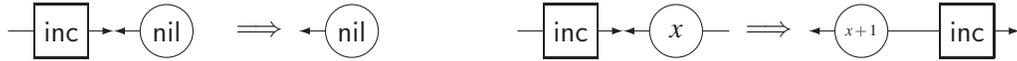
\begin{center}
\begin{mnet}{190}{20}
\putbox{10}{0}{20}{20}{$\inc$}
\putagent{60}{10}{\nil}
\putHline{0}{10}{10}
\putRvector{30}{10}{10}
\putLvector{50}{10}{10}
\puttext{95}{10}{$\Lra$}
\putagent{130}{10}{\nil}
\putLvector{120}{10}{10}
\end{mnet}
\begin{mnet}{190}{20}
\putbox{10}{0}{20}{20}{$\inc$}
\putagent{60}{10}{$x$}
\putHline{0}{10}{10}
\putRvector{30}{10}{10}
\putLvector{50}{10}{10}
\putHline{70}{10}{10}
\puttext{95}{10}{$\Lra$}
\putagent{130}{10}{\tiny $x+1$}
\putbox{160}{0}{20}{20}{$\inc$}
\putLvector{120}{10}{10}
\putHline{140}{10}{20}
\putRvector{180}{10}{10}
\end{mnet}
\end{center}

The following net rewrites using four interactions to reach a normal form:

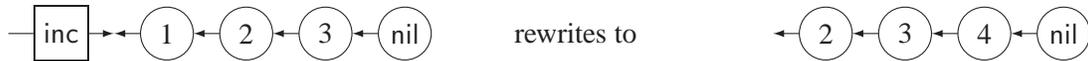
\begin{center}
\begin{mnet}{170}{20}
\putbox{20}{0}{20}{20}{$\inc$}
\putagent{70}{10}{1}
\putagent{100}{10}{2}
\putagent{130}{10}{3}
\putagent{160}{10}{\nil}
\putHline{10}{10}{10}
\putRvector{40}{10}{10}
\putLvector{60}{10}{10}
\putLvector{90}{10}{10}
\putLvector{120}{10}{10}
\putLvector{150}{10}{10}
\end{mnet}
\qquad
\begin{mnet}{30}{20}
  \puttext{30}{10}{rewrites to}
\end{mnet}
\qquad
\begin{mnet}{170}{20}
\putagent{70}{10}{2}
\putagent{100}{10}{3}
\putagent{130}{10}{4}
\putagent{160}{10}{\nil}
\putLvector{60}{10}{10}
\putLvector{90}{10}{10}
\putLvector{120}{10}{10}
\putLvector{150}{10}{10}
\end{mnet}
\end{center}
Other operations on lists/streams can be defined in a similar way.

\paragraph{Duplicating streams.}

A very important operation, because of the linearity constraint, is
the ability to explicitly copy lists and streams. With the
introduction of a new node $\delta$, we can define two interaction
rules which will copy any list or stream in the following way:

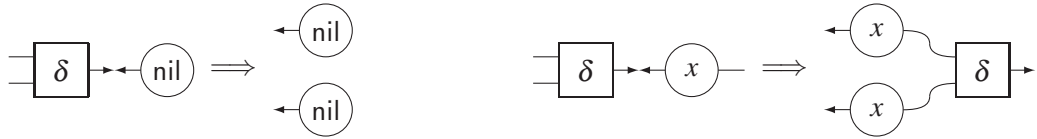
\begin{center}
\begin{mnet}{130}{60}
\putbox{10}{20}{20}{20}{$\delta$}
\putagent{60}{30}{\nil}
\putagent{120}{45}{\nil}
\putagent{120}{15}{\nil}
\putHline{0}{25}{10}
\putHline{0}{35}{10}
\putRvector{30}{30}{10}
\putLvector{50}{30}{10}
\puttext{85}{30}{$\Lra$}
\putLvector{110}{15}{10}
\putLvector{110}{45}{10}
\end{mnet}
\qquad\qquad\qquad
\begin{mnet}{190}{60}
\putbox{10}{20}{20}{20}{$\delta$}
\putagent{60}{30}{$x$}
\putagent{130}{45}{$x$}
\putagent{130}{15}{$x$}
\putbox{160}{20}{20}{20}{$\delta$}
\putHline{70}{30}{10}
\putRvector{180}{30}{10}
\qbezier(140,15)(150,15)(150,20)
\qbezier(150,20)(150,25)(160,25)
\qbezier(140,45)(150,45)(150,40)
\qbezier(150,40)(150,35)(160,35)
\putHline{0}{25}{10}
\putHline{0}{35}{10}
\putRvector{30}{30}{10}
\putLvector{50}{30}{10}
\puttext{95}{30}{$\Lra$}
\putLvector{120}{15}{10}
\putLvector{120}{45}{10}
\end{mnet}
\end{center}

We can now put both of the above systems together to compute things
like the following (that we leave as an exercise to the reader):

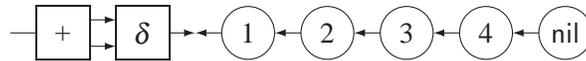
\begin{net}{220}{20}
\putbox{10}{0}{20}{20}{+}
\putbox{40}{0}{20}{20}{$\delta$}
\putagent{90}{10}{1}
\putagent{120}{10}{2}
\putagent{150}{10}{3}
\putagent{180}{10}{4}
\putagent{210}{10}{\nil}
\putHline{0}{10}{10}
\putRvector{30}{15}{10}
\putRvector{30}{5}{10}
\putRvector{60}{10}{10}
\putLvector{80}{10}{10}
\putLvector{110}{10}{10}
\putLvector{140}{10}{10}
\putLvector{170}{10}{10}
\putLvector{200}{10}{10}
\end{net}

Using these components, we can generate an infinite list of integers,
starting from a given number $n$ with the following net:

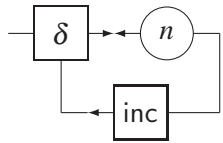
\begin{net}{80}{50}
\putbox{40}{0}{20}{20}{$\inc$}
\putbox{10}{30}{20}{20}{$\delta$}
\putagent{60}{40}{$n$}
\putHline{0}{40}{10}
\putRvector{30}{40}{10}
\putLvector{50}{40}{10}
\putHline{70}{40}{10}
\putVline{80}{10}{30}
\putVline{20}{10}{20}
\putHline{20}{10}{10}
\putLvector{40}{10}{10}
\putHline{60}{10}{20}
\end{net}

We call this cyclic net $\mathsf{ints(n)}$. There are variants of this
if we change ${\sf inc}$ to increment by 2, etc.  Remark that this net
is non-terminating. Two interactions are needed for each new number on
the output stream (one to copy the number, and another to increment one
of the copies). If we start with $n=1$, and trace the execution (after
two interactions each time) we have:

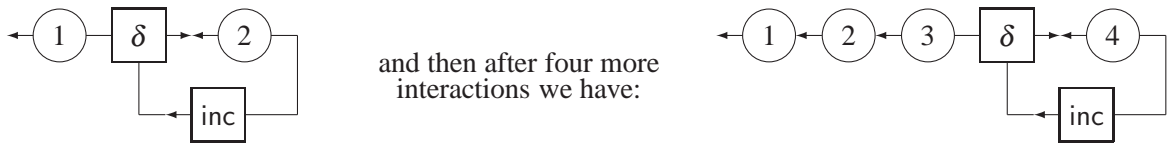
\begin{center}
  \hfill
\begin{mnet}{80}{50}
\putbox{40}{0}{20}{20}{$\inc$}
\putbox{10}{30}{20}{20}{$\delta$}
\Red{\putagent{60}{40}{$2$}}
\putHline{0}{40}{10}
\putRvector{30}{40}{10}
\putLvector{50}{40}{10}
\putHline{70}{40}{10}
\putVline{80}{10}{30}
\putVline{20}{10}{20}
\putHline{20}{10}{10}
\putLvector{40}{10}{10}
\putHline{60}{10}{20}
\Red{\putagent{-10}{40}{$1$}}
\putLvector{-20}{40}{10}
\end{mnet}
\qquad\qquad
\begin{mnet}{180}{20}
  \puttext{40}{30}{and then after four more}
  \puttext{40}{20}{interactions we have:}
\end{mnet}
\qquad
\begin{mnet}{80}{50}
\putbox{40}{0}{20}{20}{$\inc$}
\putbox{10}{30}{20}{20}{$\delta$}
\Red{\putagent{60}{40}{$4$}}
\putHline{0}{40}{10}
\putRvector{30}{40}{10}
\putLvector{50}{40}{10}
\putHline{70}{40}{10}
\putVline{80}{10}{30}
\putVline{20}{10}{20}
\putHline{20}{10}{10}
\putLvector{40}{10}{10}
\putHline{60}{10}{20}
\Red{\putagent{-10}{40}{$3$}}
\putLvector{-20}{40}{10}
\Red{\putagent{-40}{40}{$2$}}
\putLvector{-50}{40}{10}
\Red{\putagent{-70}{40}{$1$}}
\putLvector{-80}{40}{10}
\end{mnet}
\hfill
\end{center}
and so on. Thus we can generate cyclic nets, and simulate simple
data-flow with interaction nets. This net generates an infinite stream
of integers. This is an example of a process network, but uses our own
interpretation on the nets: $\delta$ and $\inc$ are considered fixed
and the numbers are interpreted as data items on a stream. In the next
section we take these ideas further.

\section{Compilation of process networks to nets}\label{sec:compiling}

Our next task is to show how to translate any KPN into an interaction
net. Not only does this highlight some connections between the
formalisms, but it also provides an implementation of process
networks. There are two possible ways to proceed:

\begin{itemize}
\item We can define a set of KPN ``combinators'', by which we mean a
  finite set of processes that can be used to represent all other
  process networks (thus all computable functions). We then just need
  to show how to translate this fixed set of processes, together with
  the communication channels and net structure, to complete the
  compilation.
  
\item We can give a general translation for each (user-programmed)
  process. This means that we need to fix the programming language
  used to define processes, and then give a translation of this
  language.

\end{itemize}

In this paper we just sketch the general translation, and then give
some example interaction systems.  We adopt a convention when writing
interaction nets (that we have followed already in the previous
examples): squares will represent processes, and circles will
represent data items on channels (streams).  The general form of the
compilation is the following:

\begin{itemize}
\item Communication channels are encoded as interaction net streams,
  as defined in Section~\ref{sec:blocks}.
  
\item Each process becomes a set of interaction net nodes that will
  simulate the functional behaviour of each process. We need to define
  the nodes and the rewrite rules for each process. Any internal state
  of a process becomes the internal state of an interaction node.
  
\item The topological structure of a process network and an interaction
  net are identical, so this structure is just copied as part of the
  compilation.
\end{itemize}

Below we give sufficient details of how process and channels are
translated so that we are able to give some examples.

\paragraph{Channels.}

Each channel in a KPN is represented as an interaction net stream. If
the channel is empty, then the stream is represented as an edge in the
interaction net. Otherwise, if $S = [x_1, \ldots , x_n]$ then we build
the following stream:

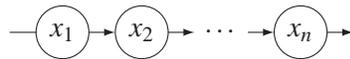
\begin{net}{150}{20}
  \putagent{20}{10}{$x_1$}
  \putagent{50}{10}{$x_2$}
\putagent{110}{10}{$x_n$}
\putHline{0}{10}{10}
\putRvector{30}{10}{10}
\putRvector{60}{10}{10}
\puttext{80}{10}{$\cdots$}
\putRvector{90}{10}{10}
\putRvector{120}{10}{10}
\end{net}

Initially, channels are usually empty at the start (so at the
compilation). However, some exceptions to this rule will be used.  In
the compilation above, if we allow other data, then we need to
translate that too. Here we just use numbers.

\paragraph{Processes.}

Each process $f$ is represented by a collection of interaction net
nodes. If $f$ has $m$ input streams and $n$ output streams:

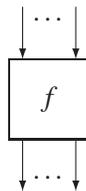
\begin{net}{40}{70}
  \putbox{10}{20}{30}{30}{$f$}
  \putDvector{15}{20}{20}
  \putDvector{35}{20}{20}
  \puttext{25}{5}{$\cdots$}
  \putDvector{15}{70}{20}
  \putDvector{35}{70}{20}
  \puttext{25}{65}{$\cdots$}
\end{net}
then this will be represented by a collection of $m$ interaction nodes
$f_1, \ldots, f_m$, where each $f_i$ has the principal port at
position $i$ corresponding to the $i$th input stream, and each with
$n+m-1$ auxiliary ports ($m-1$ input and $n$ output):
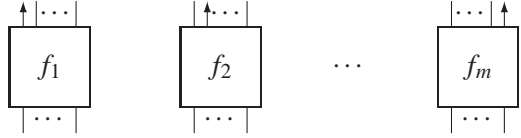
\begin{center}
\begin{mnet}{40}{55}
  \putbox{0}{10}{30}{30}{$f_1$}
  \putUvector{5}{40}{10}
  \putVline{10}{40}{10}
  \putVline{25}{40}{10}
  \putVline{5}{0}{10}
  \putVline{25}{0}{10}
  \puttext{15}{5}{$\cdots$}
  \puttext{17.5}{45}{$\cdots$}
\end{mnet}
\qquad
\begin{mnet}{40}{55}
  \putbox{0}{10}{30}{30}{$f_2$}
  \putUvector{10}{40}{10}
  \putVline{5}{40}{10}
  \putVline{25}{40}{10}
  \putVline{5}{0}{10}
  \putVline{25}{0}{10}
  \puttext{15}{5}{$\cdots$}
  \puttext{17.5}{45}{$\cdots$}
\end{mnet}
\quad
\begin{mnet}{30}{55}
    \puttext{10}{25}{$\cdots$}
\end{mnet}
\quad
\begin{mnet}{40}{50}
  \putbox{0}{10}{30}{30}{$f_m$}
  \putUvector{25}{40}{10}
  \putVline{20}{40}{10}
  \putVline{5}{40}{10}
  \putVline{5}{0}{10}
  \putVline{25}{0}{10}
  \puttext{15}{5}{$\cdots$}
  \puttext{12.5}{45}{$\cdots$}
\end{mnet}
\end{center}

For each node $f_i$ we define the corresponding interaction rules that
simulate the process.  The interaction rule $f_i$ with input on stream
$i$ becomes the process $f_j$ (where $i=j$ is a possibility) and
possibly outputs something on one of the output channels.  The
internal state is changed accordingly also.

We note that a process network uses internal state in two very
different ways: one to know which channel to read from, and the other
to store values from past inputs. Interaction nets just use the
second, because the choice of which channel to read from comes from
the different node used (which gives the position of the principal
port).  Note also that if any channel is not read by the process, then
there is no need to generate any rules for the corresponding node.

\begin{example}
  Consider a KPN that computes the running total from a channel. This
  can simply be implemented by a single process that stores internally,
  starting from 0, the accumulative sum of all the values read on the
  input channel. For each data item on the input channel, the total is
  updated, and the new total is written to the output.  Following the
  ideas of the last section, we need just one node, and one
  interaction rule to represent this system:

  \begin{net}{80}{80}
    \putbox{0}{10}{20}{20}{$s(x)$}
    \putagent{10}{60}{$y$}
    \putVline{10}{70}{10}
    \putUvector{10}{30}{10}
    \putDvector{10}{50}{10}
    \putVline{10}{0}{10}
    \puttext{40}{40}{$\Lra$}
    \putbox{60}{50}{20}{20}{\tiny $s(x+y)$}
    \putagent{70}{20}{\tiny $x+y$}
    \putDvector{70}{10}{10}
    \putUvector{70}{70}{10}
    \putVline{70}{30}{20}
  \end{net}

If we start with the node $s(0)$, we can generate the stream of
accumulating totals read so far. The process keeps a history of the
inputs (the sum of all the previous values).
\end{example}

\begin{example}
  Our next example is a variant on the previous one, where there is no
  internal state in the process. We use the network channels to keep
  the running total, which illustrates a general idea that we can
  frequently represent the internal state using additional
  channels. Consider the following process network:

  \begin{multicols}{2}
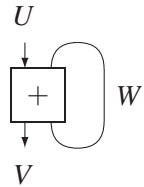
  \begin{net}{40}{90}
    \putbox{0}{20}{20}{20}{$+$}
    \putDvector{5}{20}{10}
    \putDvector{5}{50}{10}
    \qbezier(15,20)(15,10)(25,10)
    \qbezier(35,20)(35,10)(25,10)
    \putVline{35}{20}{20}
    \qbezier(35,40)(35,50)(25,50)
    \qbezier(15,40)(15,50)(25,50)
    \puttext{5}{60}{$U$}
    \puttext{5}{0}{$V$}
    \puttext{45}{30}{$W$}
  \end{net}  
\begin{verbatim}
send 0 on W
repeat
  x = wait(U)
  w = wait(W)
  send (x+w) on V
  send (x+w) on W
end
\end{verbatim}
  \end{multicols}
\end{example}

This gets compiled into the following system. Since we need to send
before read, we need to initialise the stream $W$ with $0$.  The
following is the initial net generated, and we give the two rules
corresponding to the + process:

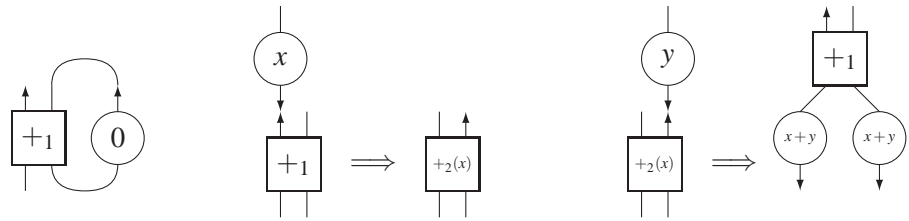
\begin{center}
\begin{mnet}{40}{60}
 \putbox{0}{20}{20}{20}{$+_1$}
 \putUvector{5}{40}{10}
 \putVline{5}{10}{10}
 \qbezier(15,20)(15,10)(27.5,10)
 \qbezier(40,20)(40,10)(27.5,10)
 \qbezier(40,50)(40,60)(27.5,60)
 \qbezier(15,50)(15,60)(27.5,60)
 \putVline{15}{40}{10}
 \putagent{40}{30}{$0$}
 \putUvector{40}{40}{10}
\end{mnet}
\qquad\qquad
\begin{mnet}{90}{80}
  \putbox{10}{10}{20}{20}{$+_1$}
  \puttext{50}{20}{$\Lra$}
  \putbox{70}{10}{20}{20}{\tiny $+_2(x)$}
  \putagent{15}{60}{$x$}
  \putVline{15}{70}{10}
  \putDvector{15}{50}{10}
  \putUvector{15}{30}{10}
  \putVline{15}{0}{10}
  \putVline{25}{0}{10}
\putVline{25}{30}{10}
\putVline{75}{0}{10}
  \putVline{85}{0}{10}
  \putVline{75}{30}{10}
  \putUvector{85}{30}{10}
\end{mnet}
\qquad\qquad
\begin{mnet}{90}{80}
  \putbox{10}{10}{20}{20}{\tiny $+_2(x)$}
  \puttext{50}{20}{$\Lra$}
  \putbox{80}{50}{20}{20}{$+_1$}
  \putagent{25}{60}{$y$}
  \putVline{25}{70}{10}
  \putDvector{25}{50}{10}
  \putUvector{25}{30}{10}
  \putVline{15}{0}{10}
  \putVline{25}{0}{10}
\putVline{15}{30}{10}
\putDvector{75}{20}{10}
  \putDvector{105}{20}{10}
  \putUvector{85}{70}{10}
  \putVline{95}{70}{10}

  \putline{75}{40}{1}{1}{10}
  \putline{105}{40}{-1}{1}{10}

  \putagent{75}{30}{\tiny $x+y$}
  \putagent{105}{30}{\tiny $x+y$}
\end{mnet}
\end{center}

We end this section by stating a very general relation between KPN
and the compiled interaction net.

\begin{theorem}[Correctness]
 If a KPN reads or writes on a channel, then the corresponding
 interaction net can make the same move.
\end{theorem}

Because we mimic the functional behaviour of the KPN as a system of
interaction nets, then this result is essentially obtained by
construction. In a longer version of this paper we give the additional
details to justify this. Finally, we remark that interaction nets are
one-step confluent, therefore we get a completeness result also: it
doesn't matter which way we evaluate a generated interaction net, it
will always output the same thing as the KPN.

\section{Representing process networks}\label{sec:examples}

Each process becomes a node in the interaction system, and we give
rules that simulate the required behaviour.  Channels become streams,
and we represent streams of data as given previously.  In this section
we give another example of the compilation, and also give some
examples of programming interaction nets directly in the style of
process networks. The alternating bit process given in Section 2 can
be represented by the following interaction net:

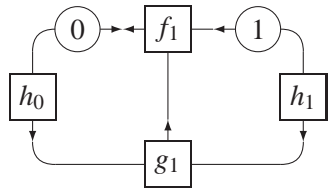
\begin{net}{140}{80}\setlength{\unitlength}{.3mm}
\putbox{60}{0}{20}{20}{$g_1$}
  \putbox{60}{60}{20}{20}{$f_1$}
  \putbox{0}{30}{20}{20}{$h_0$}
  \putbox{120}{30}{20}{20}{$h_1$}

  \putagent{30}{70}{$0$}
  \putagent{110}{70}{$1$}

  \putVline{10}{50}{10}
  \putVline{70}{30}{30}
  \putVline{130}{50}{10}

  \putHline{80}{70}{10}
  
  \putHline{20}{10}{40}
  \putHline{80}{10}{40}

\putUvector{70}{20}{10}
\putDvector{10}{30}{10}
\putDvector{130}{30}{10}

\putRvector{40}{70}{10}

\putLvector{100}{70}{10}
\putLvector{60}{70}{10}

\qbezier(10,60)(10,70)(20,70)

\qbezier(10,20)(10,10)(20,10)
\qbezier(120,10)(130,10)(130,20)
\qbezier(130,60)(130,70)(120,70)

\end{net}

\noindent
with the following six interaction rules (there are two rules for
$h_i$ that are identical):

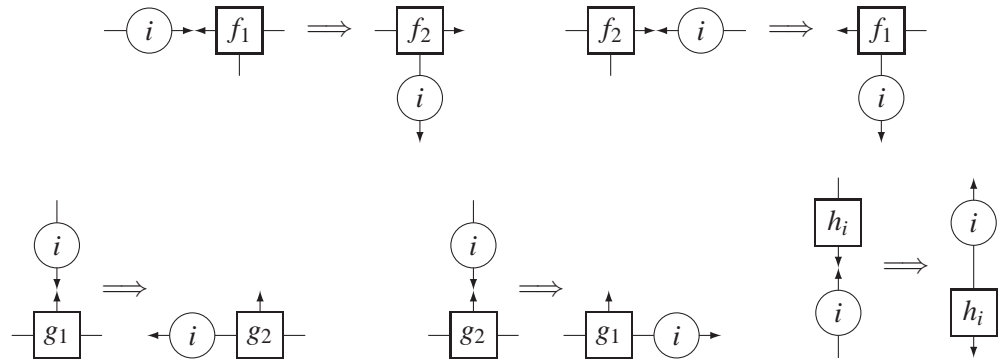
\begin{center}\setlength{\unitlength}{.3mm}
\begin{mnet}{150}{60}
  \putagent{20}{50}{$i$}
  \putbox{50}{40}{20}{20}{$f_1$}
  \putbox{130}{40}{20}{20}{$f_2$}
  \putagent{140}{20}{$i$}
  \putHline{0}{50}{10}
  \putRvector{30}{50}{10}
  \putLvector{50}{50}{10}
  \putHline{70}{50}{10}
  \putVline{60}{30}{10}
  \puttext{100}{50}{$\Lra$}
  \putHline{120}{50}{10}
  \putRvector{150}{50}{10}
  \putDvector{140}{10}{10}
  \putVline{140}{30}{10}
\end{mnet}
\qquad\qquad
\begin{mnet}{150}{60}
  \putbox{10}{40}{20}{20}{$f_2$}
  \putagent{60}{50}{$i$}
  \putbox{130}{40}{20}{20}{$f_1$}
  \putagent{140}{20}{$i$}
  \putHline{0}{50}{10}
  \putRvector{30}{50}{10}
  \putLvector{50}{50}{10}
  \putHline{70}{50}{10}
  \putVline{20}{30}{10}
  \puttext{100}{50}{$\Lra$}
  \putLvector{130}{50}{10}
  \putHline{150}{50}{10}
  \putDvector{140}{10}{10}
  \putVline{140}{30}{10}
\end{mnet}
\end{center}

\begin{center}\setlength{\unitlength}{.3mm}
\begin{mnet}{130}{50}
  \putbox{10}{0}{20}{20}{$g_1$}
  \putagent{20}{50}{$i$}
  \putHline{0}{10}{10}
  \putHline{30}{10}{10}
  \putUvector{20}{20}{10}
  \putDvector{20}{40}{10}
  \puttext{50}{30}{$\Lra$}
  \putagent{80}{10}{$i$}
  \putbox{100}{0}{20}{20}{$g_2$}
  \putLvector{70}{10}{10}
  \putHline{90}{10}{10}
  \putHline{120}{10}{10}
  \putUvector{110}{20}{10}
  \putVline{20}{60}{10}
\end{mnet}
\qquad\qquad
\begin{mnet}{130}{50}
  \putVline{20}{60}{10}
  \putbox{10}{0}{20}{20}{$g_2$}
  \putagent{20}{50}{$i$}
  \putHline{0}{10}{10}
  \putHline{30}{10}{10}
  \putUvector{20}{20}{10}
  \putDvector{20}{40}{10}
  \puttext{50}{30}{$\Lra$}
  \putbox{70}{0}{20}{20}{$g_1$}
  \putagent{110}{10}{$i$}
  \putHline{60}{10}{10}
  \putHline{90}{10}{10}
  \putRvector{120}{10}{10}
  \putUvector{80}{20}{10}
\end{mnet}
\qquad\quad
\begin{mnet}{80}{90}
  \putbox{0}{50}{20}{20}{$h_i$}
  \putbox{60}{10}{20}{20}{$h_i$}

  \putagent{10}{20}{$i$}
  \putagent{70}{60}{$i$}

  \putDvector{10}{50}{10}
  \putDvector{70}{10}{10}

  \putUvector{10}{30}{10}
  \putUvector{70}{70}{10}

  \putVline{70}{30}{20}
  \putVline{10}{70}{10}
  \putVline{10}{0}{10}
  \puttext{40}{40}{$\Lra$}
\end{mnet}
\end{center}

The choice of the principal ports directly encodes the order of
reading on channels of the original process. We next show how to write
process networks directly. Suppose we wish to build a net expressing
the following input/output behaviour that we represent as a box $f$
shown below. We can think of this as the function on pairs:
$f=\lambda\pair{x}{y}.\pair{x}{x+y})$. We can represent this using the
building blocks we have already introduced, as shown below on the
right:

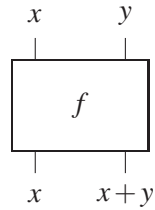
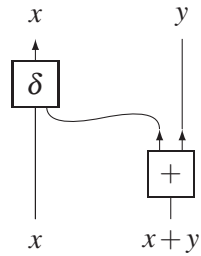
\begin{center}\setlength{\unitlength}{.3mm}
\begin{mnet}{60}{80}
\putbox{0}{20}{60}{40}{$f$}
\putVline{10}{60}{10}
\putVline{50}{60}{10}
\putVline{50}{10}{10}
\putVline{10}{10}{10}
\puttext{10}{80}{$x$}
\puttext{10}{0}{$x$}
\puttext{50}{80}{$y$}
\puttext{50}{0}{$x+y$}
\end{mnet}
\qquad\qquad\qquad\qquad
\begin{mnet}{80}{90}
\putbox{0}{60}{20}{20}{$\delta$}
\putbox{60}{20}{20}{20}{$+$}
\putUvector{10}{80}{10}
\putUvector{65}{40}{10}
\putUvector{75}{40}{10}
\putVline{75}{50}{40}
\putVline{70}{10}{10}
\putVline{10}{10}{50}
\putVline{10}{10}{50}
\qbezier(15,60)(15,50)(40,55)
\qbezier(40,55)(65,60)(65,50)
\puttext{10}{100}{$x$}
\puttext{10}{0}{$x$}
\puttext{75}{100}{$y$}
\puttext{70}{0}{$x+y$}
\end{mnet}
\end{center}
Connecting numbers to the top will give the required behaviour.  We
can iterate this function by building a feedback loop as shown below
on the left. Finally, we can put some starting values in to complete
the network, as shown below on the right:

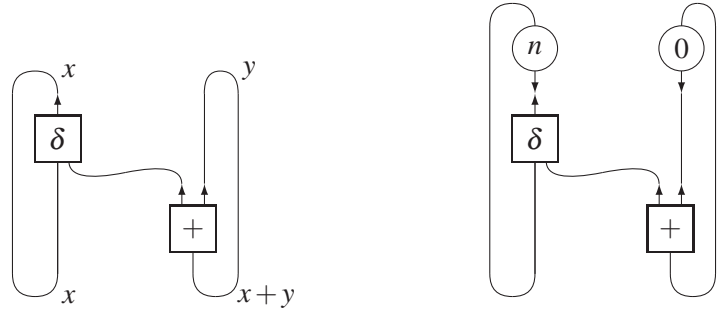
\begin{center}\setlength{\unitlength}{.3mm}
\begin{mnet}{80}{130}
\putbox{0}{60}{20}{20}{$\delta$}
\putbox{60}{20}{20}{20}{$+$}
\putUvector{10}{80}{10}
\putUvector{65}{40}{10}
\putUvector{75}{40}{10}
\putVline{75}{50}{40}
\putVline{70}{10}{10}
\putVline{10}{10}{50}
\putVline{10}{10}{50}
\qbezier(15,60)(15,50)(40,55)
\qbezier(40,55)(65,60)(65,50)
\puttext{15}{100}{$x$}
\puttext{15}{0}{$x$}
\puttext{95}{100}{$y$}
\puttext{102.5}{0}{$x+y$}
\Red{\qbezier(10,10)(10,0)(0,0)
\qbezier(0,0)(-10,0)(-10,10)
\putVline{-10}{10}{80}
\qbezier(-10,90)(-10,100)(0,100)
\qbezier(0,100)(10,100)(10,90)
\qbezier(70,10)(70,0)(80,0)
\qbezier(80,0)(90,0)(90,10)
\putVline{90}{10}{80}
\qbezier(90,90)(90,100)(82,100)
\qbezier(82,100)(75,100)(75,90)
}
\end{mnet}
\qquad\qquad\qquad\qquad\qquad
\begin{mnet}{80}{130}
\putbox{0}{60}{20}{20}{$\delta$}
\putbox{60}{20}{20}{20}{$+$}
\putUvector{10}{80}{10}
\putUvector{65}{40}{10}
\putUvector{75}{40}{10}
\putVline{75}{50}{40}
\putVline{70}{10}{10}
\putVline{10}{10}{50}
\putVline{10}{10}{50}
\qbezier(15,60)(15,50)(40,55)
\qbezier(40,55)(65,60)(65,50)
\putagent{10}{110}{$n$}
\putDvector{10}{100}{10}
\putagent{75}{110}{$0$}
\putDvector{75}{100}{10}
\qbezier(10,10)(10,0)(0,0)
\qbezier(0,0)(-10,0)(-10,10)
\putVline{-10}{10}{110}
\qbezier(-10,120)(-10,130)(0,130)
\qbezier(0,130)(10,130)(10,120)
\qbezier(70,10)(70,0)(80,0)
\qbezier(80,0)(90,0)(90,10)
\putVline{90}{10}{110}
\qbezier(90,120)(90,130)(82,130)
\qbezier(82,130)(75,130)(75,120)
\end{mnet}
\end{center}

If we look at the output of the $+$ node, we will get $n$, $2n$, $3n$,
etc. This network will compute a stream of multiples of $n$. We can
also control the computation by restricting the iterations by
introducing a counter. To do this we need to know when to stop, and
how to extract the result.  We can do this by adding in some
synchronisation that can be encoded with standard ideas from
interaction nets.

We give two final examples to illustrate how easily different nets can
be built. These networks compute a stream of factorial numbers and
a stream of Fibonacci numbers respectively:

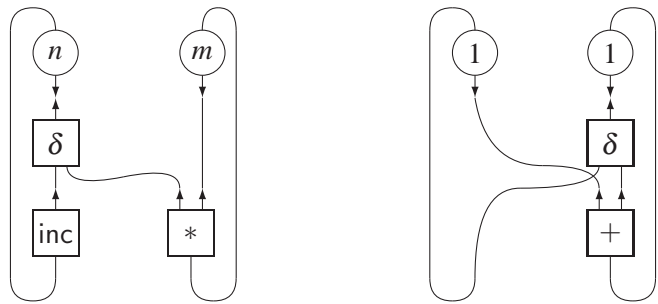
\begin{center}\setlength{\unitlength}{.3mm}
\begin{mnet}{80}{130}
\putbox{0}{60}{20}{20}{$\delta$}
\putbox{60}{20}{20}{20}{$*$}
\putUvector{10}{80}{10}
\putUvector{65}{40}{10}
\putUvector{75}{40}{10}
\putVline{75}{50}{40}
\putVline{70}{10}{10}
\putVline{10}{10}{10}
\putVline{10}{50}{10}
\putUvector{10}{40}{10}
\putbox{0}{20}{20}{20}{$\inc$}
\qbezier(15,60)(15,50)(40,55)
\qbezier(40,55)(65,60)(65,50)
\putagent{10}{110}{$n$}
\putDvector{10}{100}{10}
\putagent{75}{110}{$m$}
\putDvector{75}{100}{10}
\qbezier(10,10)(10,0)(0,0)
\qbezier(0,0)(-10,0)(-10,10)
\putVline{-10}{10}{110}
\qbezier(-10,120)(-10,130)(0,130)
\qbezier(0,130)(10,130)(10,120)
\qbezier(70,10)(70,0)(80,0)
\qbezier(80,0)(90,0)(90,10)
\putVline{90}{10}{110}
\qbezier(90,120)(90,130)(82,130)
\qbezier(82,130)(75,130)(75,120)
\end{mnet}
\qquad\qquad\qquad\qquad
\begin{mnet}{80}{120}
\putbox{60}{60}{20}{20}{$\delta$}
\putbox{60}{20}{20}{20}{$+$}
\putUvector{70}{80}{10}
\putUvector{65}{40}{10}
\putUvector{75}{40}{10}
\putVline{75}{50}{10}
\putVline{70}{10}{10}
\qbezier(10,90)(15,60)(40,60)
\qbezier(40,60)(65,60)(65,50)
\putagent{10}{110}{$1$}
\putDvector{10}{100}{10}
\putagent{70}{110}{$1$}
\putDvector{70}{100}{10}
\qbezier(10,10)(10,0)(0,0)
\qbezier(0,0)(-10,0)(-10,10)
\putVline{-10}{10}{110}
\qbezier(-10,120)(-10,130)(0,130)
\qbezier(0,130)(10,130)(10,120)
\qbezier(70,10)(70,0)(80,0)
\qbezier(80,0)(90,0)(90,10)
\putVline{90}{10}{110}
\qbezier(90,120)(90,130)(80,130)
\qbezier(80,130)(70,130)(70,120)
\qbezier(65,60)(65,50)(30,50)
\qbezier(10,10)(10,50)(30,50)
\end{mnet}
\end{center}

\section{Conclusion}\label{sec:conc}

Interaction nets give a very easy way of implementing data-flow
networks. There are several parallel implementations of interaction
nets now available, so this means that we can also take advantage of
the natural parallelism that arises by writing programs in this
way. Programming algorithms in interaction nets by encoding a process
network can give a very efficient encoding of the algorithm: the
example above for Fibonacci needs just three kinds of nodes and two
rewrite rules to generate an infinite stream of Fibonacci numbers
which is simpler and more efficient than any other way.

There are interesting extensions to the ideas presented here that are
currently being investigated. For instance, is is possible to allow
KPN to dynamically create new processes. This fits very naturally with
interaction nets. Notions of second order networks (see for instance
\cite{Matthews}) have been discussed, and interesting relations again
arise with past work on interaction nets. We hope to report on some of
these ideas in a longer version of this paper.

\bibliographystyle{eptcs} 
\bibliography{bibfile}

\end{document}